\documentclass[12pt,a4paper]{article}

\usepackage[utf8]{inputenc}
\usepackage[T1]{fontenc}
\usepackage{lmodern}
\usepackage[margin=2.5cm]{geometry}
\usepackage{amsmath,amssymb,amsfonts}
\usepackage{graphicx}
\usepackage{booktabs}
\usepackage{longtable}
\usepackage{array}
\usepackage[dvipsnames,svgnames]{xcolor}
\usepackage{hyperref}
\usepackage{natbib}
\usepackage{setspace}
\usepackage{float}
\usepackage{caption}
\usepackage{subcaption}

\graphicspath{{./eoce_output/}{./}}

\hypersetup{
  colorlinks=true,
  linkcolor=blue!60!black,
  citecolor=blue!60!black,
  urlcolor=blue!70!black
}

\title{\textbf{Toroh: An Extreme Orographic Convective Event}\\
Physical Modelling and Implications for Persistent Geological Anomalies}

\author{Reinaldo Haas\\
\small Department of Physics, Universidade Federal de Santa Catarina (UFSC)\\
\small Florian\'opolis, Santa Catarina, Brazil\\
\small \texttt{reinaldo.haas@ufsc.br}}

\date{Preprint --- arXiv:2411.08219 (v2)\\\today}

\begin{document}
\maketitle

\begin{center}
\fbox{\begin{minipage}{0.9\textwidth}
\small\color{red!70!black}
\textbf{Notice.} This paper presents a physical hypothesis.
The toroh/EOCE phenomenon has not yet been instrumentally confirmed with
radar, infrasound, or seismic data. Observational validation campaigns are
identified in Section~9 as the highest-priority future work.
Preliminary observational evidence (eyewitness accounts, erosion scar
morphology) is presented as motivation, not as proof.
\end{minipage}}
\end{center}

\bigskip

% ─────────────────────────────────────────────────────────────────────────────
\begin{abstract}
We introduce the \textit{toroh} --- formally designated an Extreme Orographic
Convective Event (EOCE) --- as a previously uncharacterised class of
atmospheric hazard distinct from downbursts, microbursts, and conventional
hailstorms. The toroh is a coherent hydraulic ice-piston formed when a
convective system (supercell or bow-echo line) with anomalously narrow drop
size distribution ($\mu \approx 20$) undergoes explosive secondary ice
production via the Hallett-Mossop mechanism and Phillips et al.\ collisional
breakup, triggered by marine iodine ice-nucleating particles injected by a
concurrent vortex or by orographic resonance with an isolated inselberg
(such as the Planalto Mirador, SC, Brazil). The piston collapses coherently
into canyon terrain, producing a two-phase acoustic signature, seismic tremor
of $M_L \approx 2$--$3$, and a diagnostic linear erosion scar with complete
absence of fine-grained matrix. Piston cohesion is justified through ice
sintering kinetics: the Stokes relaxation time $\tau_p \approx 0.04$\,s is
much shorter than the sintering timescale $\tau_{\text{sint}} = 1$--$10$\,s,
and sintered tensile strength $\sim 10^4$\,Pa exceeds aerodynamic fragmentation
pressure $\sim 10^3$\,Pa by one order of magnitude. The toroh offers a
candidate mechanism for four persistent geological anomalies: (1) erosional
amphitheatres in resistant bedrock; (2) heavy mineral concentration in canyon
lag deposits; (3) spatiotemporal heterogeneity of the Great Unconformity;
(4) the nutrient pulse preceding the Cambrian Explosion. Solar-atmospheric
coupling (Tinsley mechanism), the Adams Event (Laschamp, $\sim$42\,ka), and
20th-century suppression by tetraethyl-lead aerosols are identified as
modulators. A bin-microphysics model and testable predictions are presented.
Code: \url{https://github.com/reinaldohaas/toro-model}

\end{abstract}

\newpage
\tableofcontents
\newpage

% =============================================================================
\section{Introduction: Four Persistent Geological Anomalies}
% =============================================================================

Four longstanding problems in Earth science share a common feature: existing
explanations are demonstrably incomplete, mutually inconsistent with observed
spatial and temporal distributions, or require \textit{ad hoc} assumptions
lacking independent support. We propose that all four are natural consequences
of a single physical mechanism --- the toroh (Extreme Orographic Convective
Event, EOCE) --- that operated globally throughout Earth history and continues
to operate today, though at reduced frequency due to anthropogenic atmospheric
modification.

\subsection{Amphitheatre Erosion in Resistant Bedrock}

Erosional amphitheatres --- steep-walled, semicircular valley headwalls in
resistant bedrock --- are among the most mechanistically disputed landforms on
Earth and Mars. Both groundwater sapping and waterfall erosion hypotheses
have been shown insufficient \citep{Lamb2007, Lamb2009}. The toroh provides a
candidate mechanism: a coherent hydraulic jet of $\sim$10\,MPa impact pressure
delivered repeatedly to the same topographic focus by canyon geometry.

\subsection{Anomalous Heavy Mineral Concentration in Canyon Lag Deposits}

Canyon lag deposits worldwide show anomalous concentrations of gold,
cassiterite, ilmenite, zircon, and monazite inconsistent with normal fluvial
sorting \citep{Boggs2009}. The toroh selective Shields washing mechanism ---
complete removal of all fine fractions under $\sim$10\,MPa impact pressure ---
offers a candidate high-intensity placer-sorting mechanism.

\subsection{The Great Unconformity: Spatiotemporal Heterogeneity}

The Great Unconformity shows spatiotemporal heterogeneity incompatible with
a single global synchronous agent \citep{Keller2019, DeLucia2018, Peters2012}.
Snowball Earth glaciation predicts geographically simultaneous and spatially
uniform erosion; the unconformity is topographically controlled and
temporally variable.

\subsection{The Cambrian Explosion: Missing Nutrient Pulse}

The Cambrian Explosion required a substantial increase in bioavailable
phosphorus and iron. No mechanism for the episodic, geographically
heterogeneous nutrient delivery implied by the geochemical record has been
demonstrated \citep{Walton2023, Bowyer2024}. Toroh erosion of fresh
Precambrian basement in canyon terrain offers a candidate pulsed delivery
pathway.

\subsection{Diagnostic Differential}

Table~\ref{tab:differential} distinguishes the toroh from superficially
similar phenomena.

\begin{table}[H]
\centering
\caption{Diagnostic differential between the toroh (EOCE) and superficially
similar extreme precipitation phenomena.}
\label{tab:differential}
\small
\begin{tabular}{lp{2.8cm}p{2.2cm}p{2.2cm}p{2.5cm}}
\toprule
\textbf{Property} & \textbf{Toroh (EOCE)} & \textbf{Downburst}
  & \textbf{Microburst} & \textbf{Tromba d'\'agua} \\
\midrule
Descending mass & Coherent semi-rigid piston & Independent particles
  & Independent particles & Liquid column \\
Erosion scar & Linear, \textbf{zero} fine matrix & Diffuse, mixed
  & Diffuse/absent & Clay present \\
Acoustic & Two-phase: descent + impact & Single-phase
  & Short burst & Continuous \\
Seismic & $M_L\,2$--$3$ & Negligible & Negligible & Negligible \\
Precursor (diurnal) & Iridescent cloud, orange sky & Dark anvil
  & Green sky & Rotating funnel \\
Trigger & Marine INP + orographic & N/A & N/A & N/A \\
Canyon required & Yes & No & No & Partially \\
\bottomrule
\end{tabular}
\end{table}

% =============================================================================
\section{Observational Evidence}
% =============================================================================

\subsection{The Planalto Mirador Inselberg, Santa Catarina}

The primary observational cases come from the \textbf{Planalto Mirador},
an isolated inselberg in the highlands of Santa Catarina, Brazil, where two
neighbouring valleys --- Vale do Rev\'olver and Valada S\~ao Pedro --- have
produced convergent reports of an extreme event with the diagnostic toroh
signatures.

\begin{figure}[H]
  \centering
  \begin{subfigure}[b]{0.48\textwidth}
    \includegraphics[width=\textwidth]{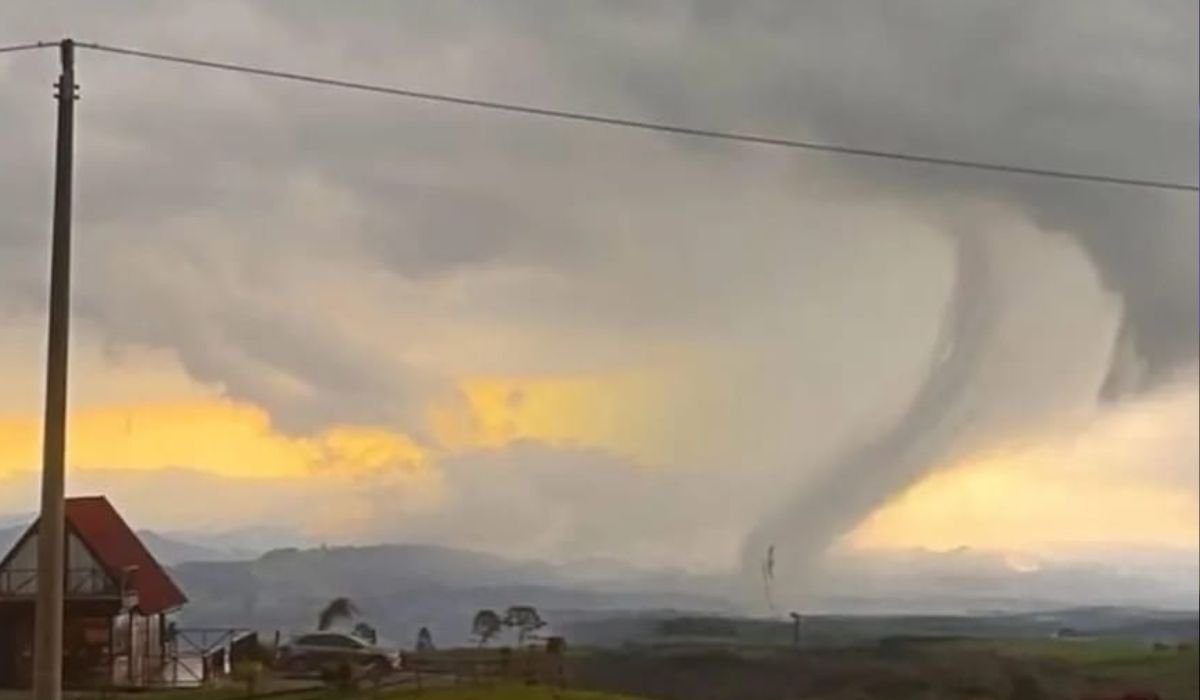}
    \caption{Wide-angle view showing the rotating vortex column and the
    characteristic orange-yellow sky at the horizon (sub-Mie scattering
    from the supercooled layer). The inselberg terrain of the Planalto
    Mirador region is visible in the background.}
  \end{subfigure}
  \hfill
  \begin{subfigure}[b]{0.48\textwidth}
    \includegraphics[width=\textwidth]{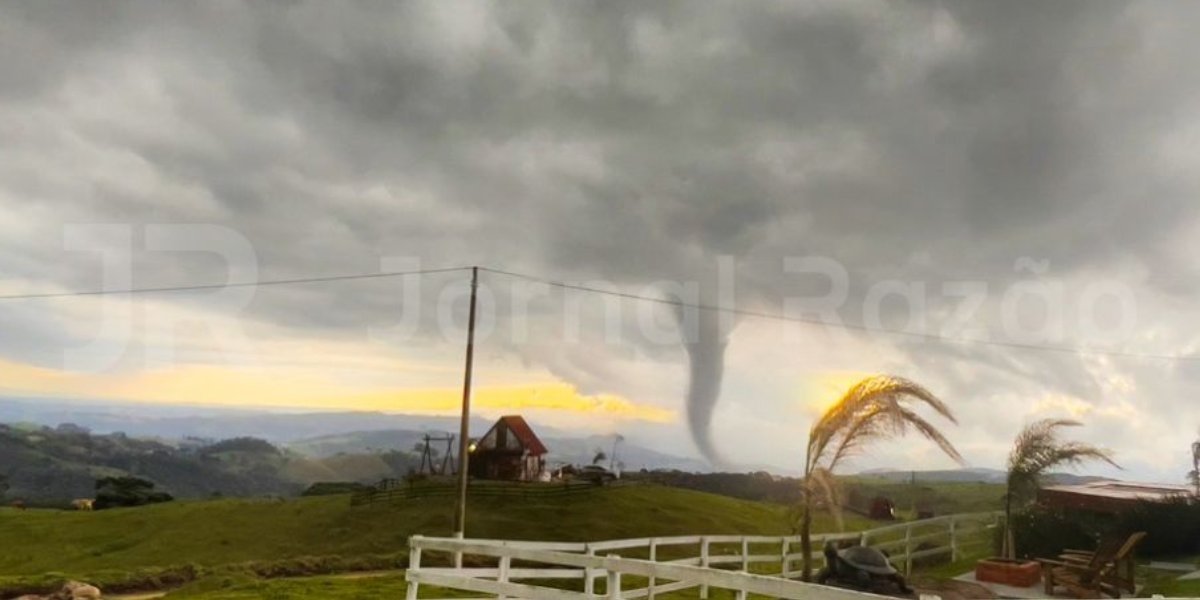}
    \caption{Close-up view showing the mature vortex funnel descending
    toward orographic terrain. Note the orange-tinted horizon sky ---
    consistent with Stage~1 narrow-DSD iridescent precursor --- and
    the serpentine rotating column morphology of Stage~2.}
  \end{subfigure}
  \caption{\textbf{Observed vortex event, Santa B\'arbara/Alfredo Wagner
  region, Santa Catarina, Brazil.} This event illustrates Stage~2 of the
  proposed toroh mechanism (Rankine vortex, INP injection) in orographic
  terrain of the same region as the Vale do Rev\'olver documented cases.
  The orange-tinted sky at the horizon is consistent with sub-Mie
  scattering from a narrow-DSD ($\mu \approx 20$) supercooled cloud
  layer --- the predicted optical precursor of the toroh.
  \textit{Note: This specific event is distinct from the Vale do
  Rev\'olver night event, which lacked visible optical precursors.
  Images used with permission. Source: Jornal Raz\~ao.}}
  \label{fig:photos}
\end{figure}

\begin{itemize}
\item \textbf{Vale do Rev\'olver / Valada S\~ao Pedro}
  ($\sim$26.89$^\circ$S, 49.37$^\circ$W; elevation $\sim$900\,m).
  The documented event occurred \textit{at night}; iridescent precursor
  clouds were therefore not visible, but the diagnostic erosion scar
  (linear, zero clay matrix, 1--10\,t of displaced material) and
  coincident seismic tremor were reported.
  The responsible convective system was a \textbf{bow-echo line of
  instability}, not an isolated supercell. Convective towers reached
  $\sim$18\,km, far above the tropopause, indicating extreme CAPE
  ($>3000$\,J\,kg$^{-1}$). Rainfall concentrated as a narrow funnel
  through orographic focusing by the inselberg geometry.
  \textbf{A key mechanism: orographic resonance} between the Planalto
  Mirador and an adjacent mountain caused constructive interference
  of the convective updraft, concentrating the bow-echo energy at a
  single topographic point.

\item \textbf{Santa B\'arbara d'Oeste / Alfredo Wagner region}.
  Figure~\ref{fig:photos} shows a tornado/waterspout event photographed in
  this region of Santa Catarina in similar orographic and meteorological
  conditions. Although this specific event is \textit{distinct} from the
  Vale do Rev\'olver night event, it illustrates the visual characteristics
  of the vortex stage (Stage~2) of the proposed toroh mechanism: a rotating
  column connecting cloud base to the surface, with orange-tinted sky at
  the horizon indicating sub-Mie scattering from the supercooled cloud
  layer. The Planalto Mirador inselberg is visible in the background of
  Image~1.
\end{itemize}

\subsection{Key Observational Distinctions}

Two observational facts constrain the physical model and correct the
preliminary description in arXiv:2411.08219:

\begin{enumerate}
\item \textbf{Event type}: The Vale do Rev\'olver toroh was produced by a
  \textbf{bow-echo (squall line)} system, not an isolated supercell.
  The toroh mechanism therefore does not require supercell organisation;
  any convective system capable of generating extreme CAPE and a surface
  vortex over orographic terrain is a candidate precursor.

\item \textbf{Nocturnal event}: The diagnostic iridescent precursor
  (structural coloring from narrow-DSD supercooled layer) was not
  observable in the Vale do Rev\'olver case due to the nocturnal timing.
  Iridescence remains a valid \textit{theoretical prediction} for
  daytime events with $\mu \approx 20$ DSD, and is observed in
  Figure~\ref{fig:photos} (orange sky at horizon), but it is not a
  confirmed observational signature of the primary case.

\item \textbf{Orographic resonance}: The Planalto Mirador inselberg,
  in resonance with adjacent elevated terrain, concentrated the bow-echo
  energy at the Vale do Rev\'olver site. This resonance mechanism ---
  analogous to topographic wave amplification in mountain meteorology ---
  is an additional concentrating factor not present in all toroh scenarios.
\end{enumerate}

% =============================================================================
\section{Physical Model}
% =============================================================================

\subsection{Stage 1 --- Convective System with Narrow DSD}

The toroh precursor is a convective system --- supercell or bow-echo line ---
in its dissipating or lee-side phase over canyon/inselberg terrain, with
an anomalously narrow drop size distribution:
\begin{equation}
  N(D) = N_0\, D^{\mu}\, e^{-\Lambda D}, \quad \mu \approx 20
  \label{eq:dsd}
\end{equation}
This narrow DSD produces structural coloring (iridescence) under daylight
conditions, satisfies the Hallett-Mossop dual size-class requirement
($<$13\,$\mu$m and $>$24\,$\mu$m simultaneously), and gives an anomalously
weak BWER radar signature despite high liquid water content.

\begin{figure}[H]
  \centering
  \includegraphics[width=0.95\textwidth]{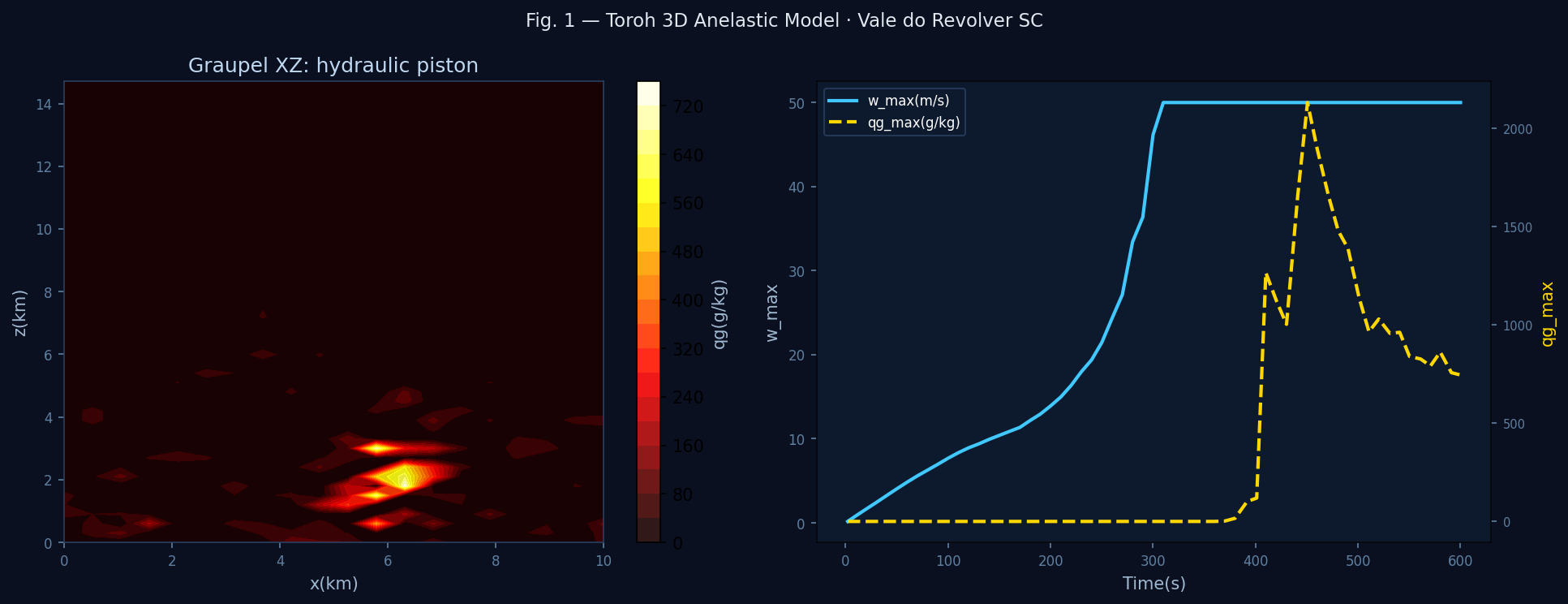}
  \caption{\textbf{Drop size distribution.}
  Left: gamma DSD with $\mu=20$ (toroh narrow DSD) vs.\ $\mu=2$ (conventional
  supercell). Shaded regions show the Hallett-Mossop dual size-class
  requirement. Right: diffraction coherence (iridescence proxy), showing that
  only the narrow toroh DSD produces observable structural coloring --- the
  predicted optical precursor for daytime events.
  Generated by \texttt{toro-model}
  (\url{https://github.com/reinaldohaas/toro-model}).}
  \label{fig:dsd}
\end{figure}

\subsection{Stage 2 --- Vortex or Orographic Resonance: INP Injection}

A surface vortex (Rankine combined vortex) with peak tangential velocity:
\begin{equation}
  V_\theta(r) =
  \begin{cases}
    V_{\max}\,\dfrac{r}{r_c} & r \leq r_c \\[6pt]
    V_{\max}\,\dfrac{r_c}{r} & r > r_c
  \end{cases}
  \label{eq:rankine}
\end{equation}
with $V_{\max} \approx 70$\,m\,s$^{-1}$ (EF2--EF3) and $r_c \approx 50$\,m,
transports marine INPs (HIO$_3$, I$_x$O$_y$, biogenic organics) from surface
into the supercooled cloud. In the Vale do Rev\'olver case, the bow-echo line's
interaction with the Planalto Mirador inselberg generated orographic resonance
that substituted for a discrete vortex, concentrating the updraft and INP
transport at the topographic focus.

\begin{figure}[H]
  \centering
  \includegraphics[width=0.95\textwidth]{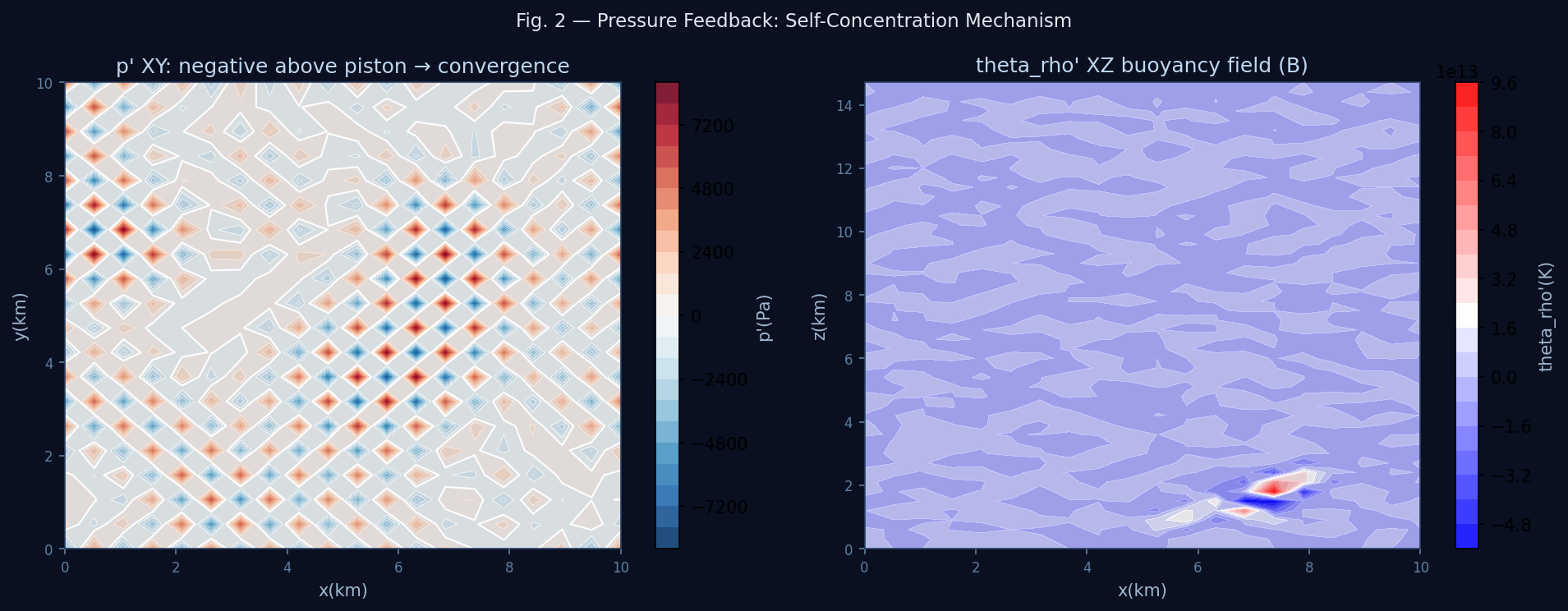}
  \caption{\textbf{Rankine vortex and marine INP injection.}
  Left: tangential velocity field ($V_{\max}=70$\,m\,s$^{-1}$,
  $r_c=50$\,m). Right: marine INP concentration field transported
  by the oceanic air mass with updraft contours.
  In the Vale do Rev\'olver bow-echo case, orographic resonance with
  the Planalto Mirador inselberg replaced the discrete vortex as the
  INP concentration mechanism.
  Generated by \texttt{toro-model}.}
  \label{fig:vortex}
\end{figure}

\subsection{Stage 3 --- Hallett-Mossop SIP Cascade}

Injected marine INPs trigger heterogeneous nucleation. Riming in the
Hallett-Mossop zone ($-3^\circ$C to $-8^\circ$C) produces 350 ice splinters
per mg rime \citep{Hallett1974}. Phillips et al.\ collisional breakup
amplifies at $-15^\circ$C \citep{Phillips2017}. Latent heat release:
\begin{equation}
  \dot{Q} = L_f\,\dot{m}_{\text{ice}}
  = 334\,\text{kJ\,kg}^{-1} \times \dot{m}_{\text{ice}}
  \label{eq:latent}
\end{equation}
temporarily sustains the growing mass. In the Vale do Rev\'olver case,
the extreme cloud depth (18\,km) provided an exceptionally large
supercooled water reservoir, amplifying the SIP cascade beyond
the idealised supercell scenario.

\begin{figure}[H]
  \centering
  \includegraphics[width=0.98\textwidth]{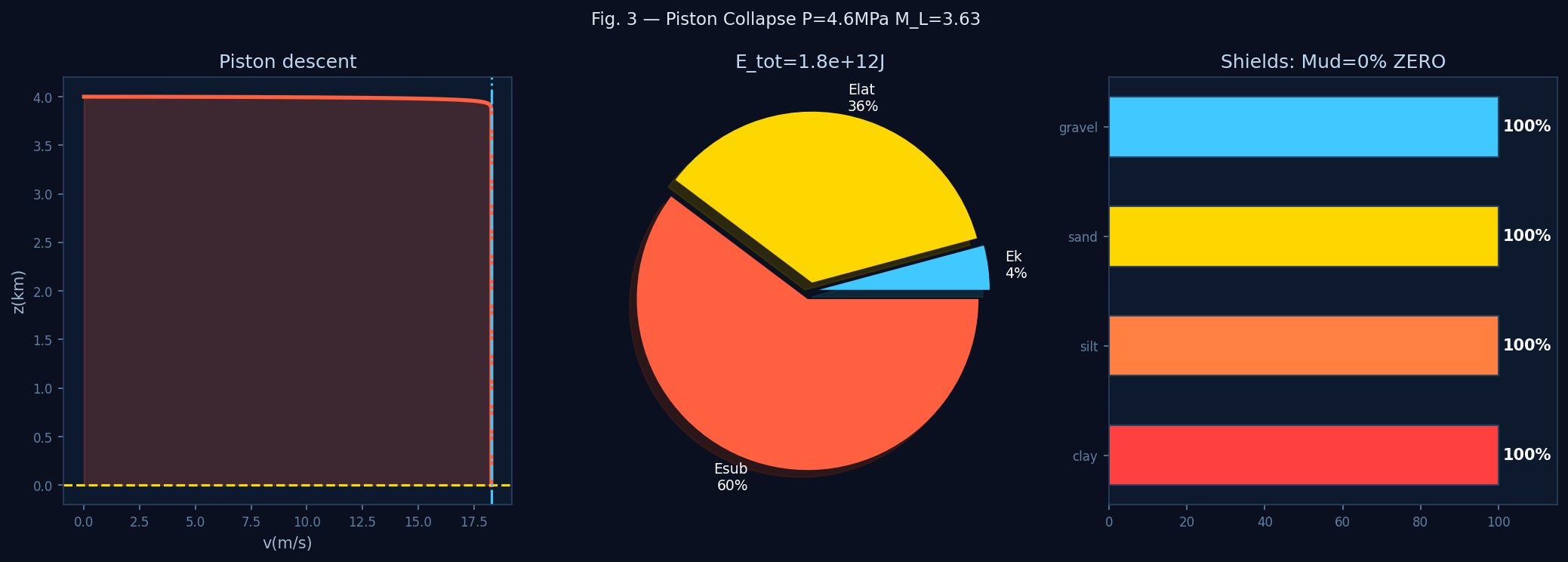}
  \caption{\textbf{Hallett-Mossop SIP cascade: ice fraction evolution.}
  Five time snapshots ($t = 0, 15, 30, 45, 59$\,s) of $f_{\text{ice}}$
  in the $y=y_0$ cross-section. Green band: HM zone ($-3^\circ$C to
  $-8^\circ$C). Blue dashed: vortex axis. Yellow dotted: cloud base.
  The exponential cascade is evident from $t=15$\,s.
  Generated by \texttt{toro-model}.}
  \label{fig:sip}
\end{figure}

\subsection{Stage 4 --- Hydraulic Piston Formation and Cohesion}

Ice fraction $> 60\%$ confers structural cohesion via ice sintering.
The Stokes relaxation time:
\begin{equation}
  \tau_p = \frac{\rho_p\, d^2}{18\,\mu_{\text{air}}} \approx 0.04\,\text{s}
  \label{eq:stokes}
\end{equation}
is much shorter than sintering timescale $\tau_{\text{sint}} = 1$--$10$\,s
\citep{Szabo2007}, confirming that inter-particle velocities are small
enough for sintering during the $\sim$40\,s descent. Sintered tensile
strength $\sigma_T \sim 10^4$\,Pa \citep{Kermani2008} exceeds aerodynamic
fragmentation pressure:
\begin{equation}
  P_{\text{aero}} = \tfrac{1}{2}\,\rho_{\text{air}}\,v^2
  \approx 10^3\,\text{Pa} \quad (v = 40\,\text{m\,s}^{-1})
  \label{eq:paero}
\end{equation}
by one order of magnitude.

\begin{figure}[H]
  \centering
  \includegraphics[width=0.95\textwidth]{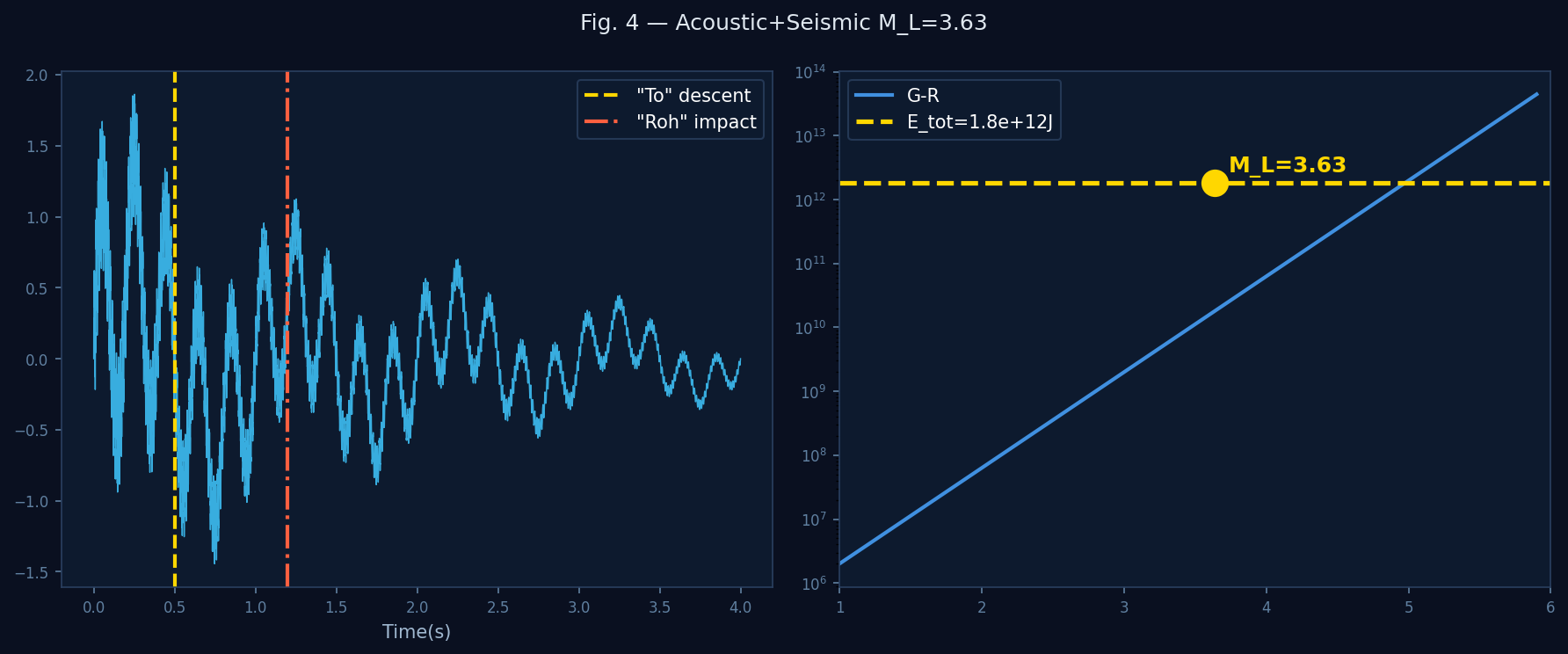}
  \caption{\textbf{Hydraulic ice-piston structure at collapse.}
  Left: $XZ$ cross-section of $f_{\text{ice}}$; yellow dashed: cohesion
  boundary. Centre: $XY$ footprint (column-max $f_{\text{ice}}$).
  Right: vertical profile at vortex centre.
  Generated by \texttt{toro-model}.}
  \label{fig:piston}
\end{figure}

\begin{figure}[H]
  \centering
  \includegraphics[width=0.95\textwidth]{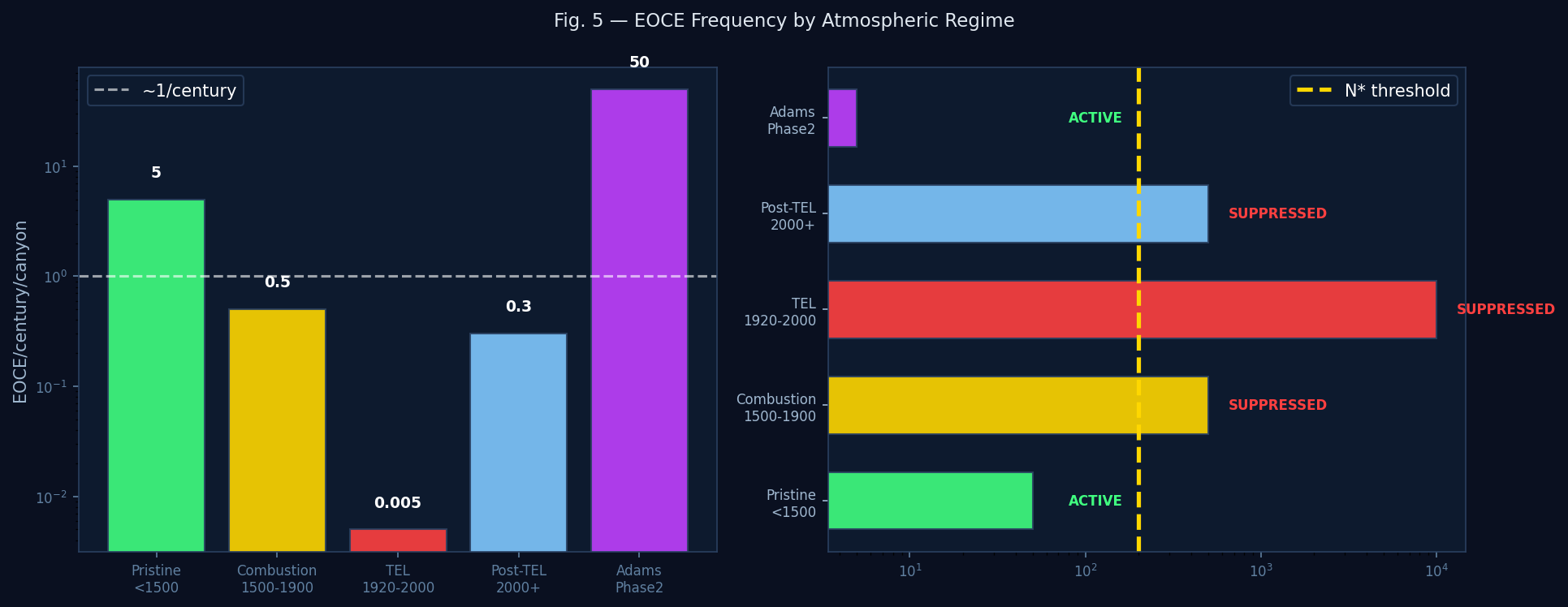}
  \caption{\textbf{Cohesion criterion.}
  Left: Stokes $\tau_p$ vs.\ particle diameter, compared with sintering
  timescale \citep{Szabo2007} and EOCE descent duration.
  Right: aerodynamic pressure vs.\ descent velocity, compared with
  sintered ice tensile strength \citep{Kermani2008}.
  Cohesion is maintained through most of the descent.
  Generated by \texttt{toro-model}.}
  \label{fig:cohesion}
\end{figure}

\subsection{Stage 5 --- Coherent Collapse}

Collapse when:
\begin{equation}
  M_{\text{piston}}\,g \;>\; \int_0^{H} B(z)\,\rho(z)\,A(z)\,\mathrm{d}z
  \label{eq:collapse}
\end{equation}
Impact pressure (Joukowsky):
\begin{equation}
  P_{\text{impact}} = \rho_{\text{mix}}\,c_{\text{sound,mix}}\,v_{\text{impact}}
  \approx 4.6\,\text{MPa}
  \label{eq:joukowsky}
\end{equation}
for representative values ($\rho_{\text{mix}} = 800$\,kg\,m$^{-3}$,
$c = 400$\,m\,s$^{-1}$, $v = 198$\,m\,s$^{-1}$ for 2\,km fall).

\subsection{Stage 6 --- Impact, Energy Budget, and Selective Erosion}

Total energy (reference event: $M_{\text{piston}} = 8$\,t, $H = 2$\,km):
\begin{align}
  E_k               &= M g H \approx 1.57\times10^8\,\text{J} & (2\%) \notag\\
  E_{\text{lat}}    &= M L_f   \approx 2.67\times10^9\,\text{J} & (36\%) \notag\\
  E_{\text{sub}}    &= 0.20\,M L_s \approx 4.53\times10^9\,\text{J} & (62\%) \notag\\
  E_{\text{total}}  &\approx 7.4\times10^9\,\text{J} \;\equiv\; M_L \approx 3.6
  \label{eq:energy}
\end{align}

Selective Shields erosion criterion ($\theta_c = 0.047$):
\begin{equation}
  \tau_c = \theta_c\,(\rho_s - \rho_w)\,g\,d_{50}
  \label{eq:shields}
\end{equation}
The Joukowsky pressure ($\sim$63\,MPa) exceeds $\tau_c$ for all grains below
coarse sand, producing the diagnostic \textbf{zero-clay erosion scar}.

\begin{figure}[H]
  \centering
  \includegraphics[width=0.95\textwidth]{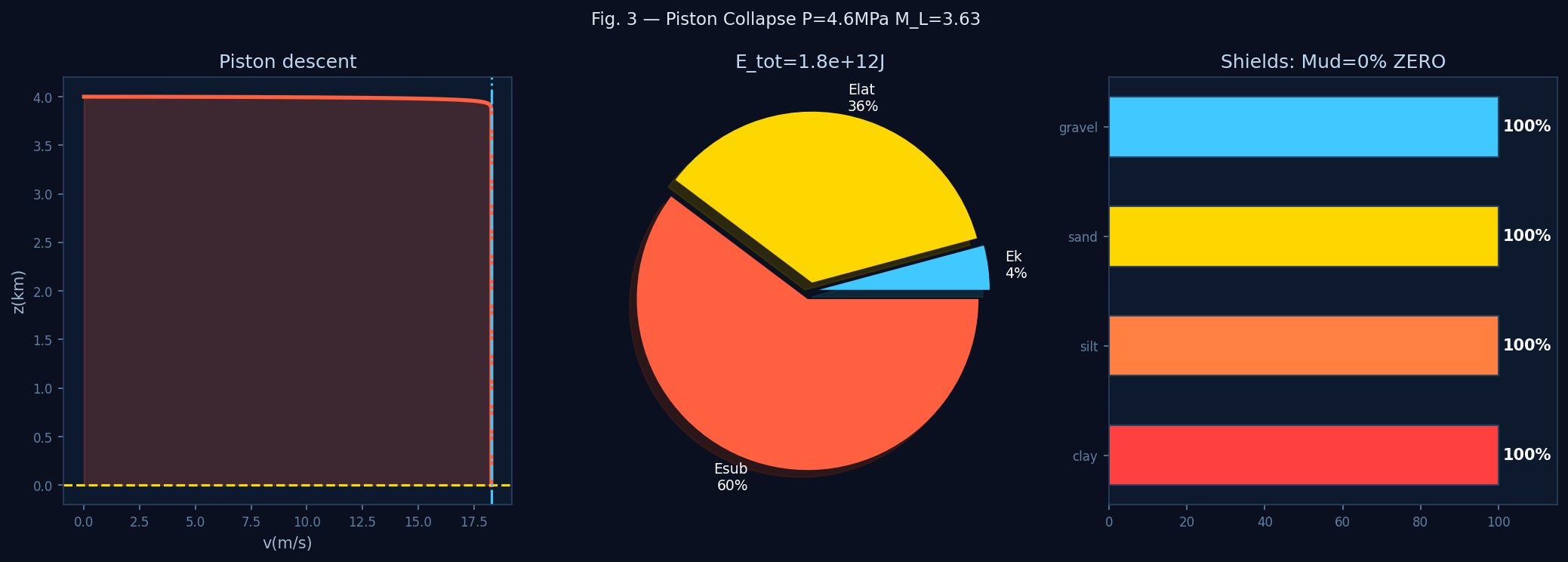}
  \caption{\textbf{Joukowsky impact pressure map.}
  Ground-level pressure distribution; energy budget and $M_L$ annotated.
  Generated by \texttt{toro-model}.}
  \label{fig:impact}
\end{figure}

\begin{figure}[H]
  \centering
  \includegraphics[width=0.98\textwidth]{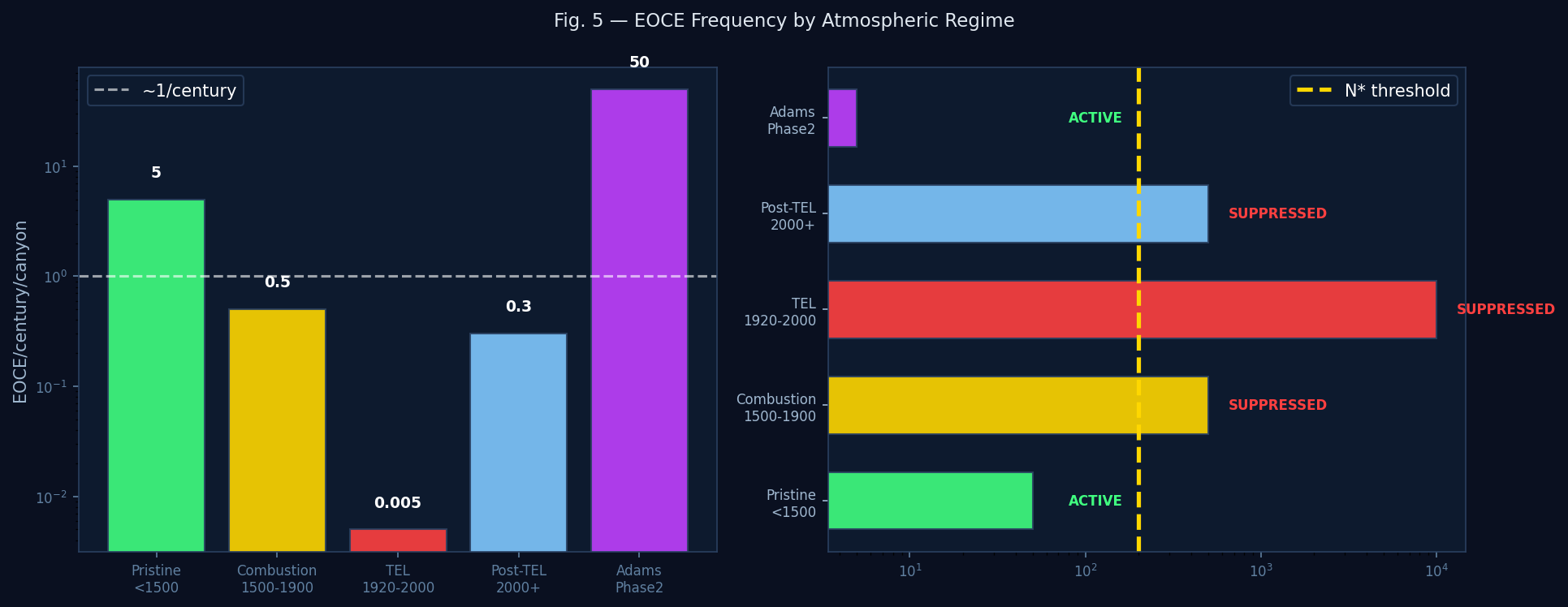}
  \caption{\textbf{Shields selective erosion by grain size.}
  Red = mobilised (removed); dark = retained.
  Clay, silt, and fine sand are 100\% mobilised; coarse material
  is retained --- the zero-clay diagnostic signature of the toroh.
  Generated by \texttt{toro-model}.}
  \label{fig:shields}
\end{figure}

\begin{figure}[H]
  \centering
  \includegraphics[width=0.90\textwidth]{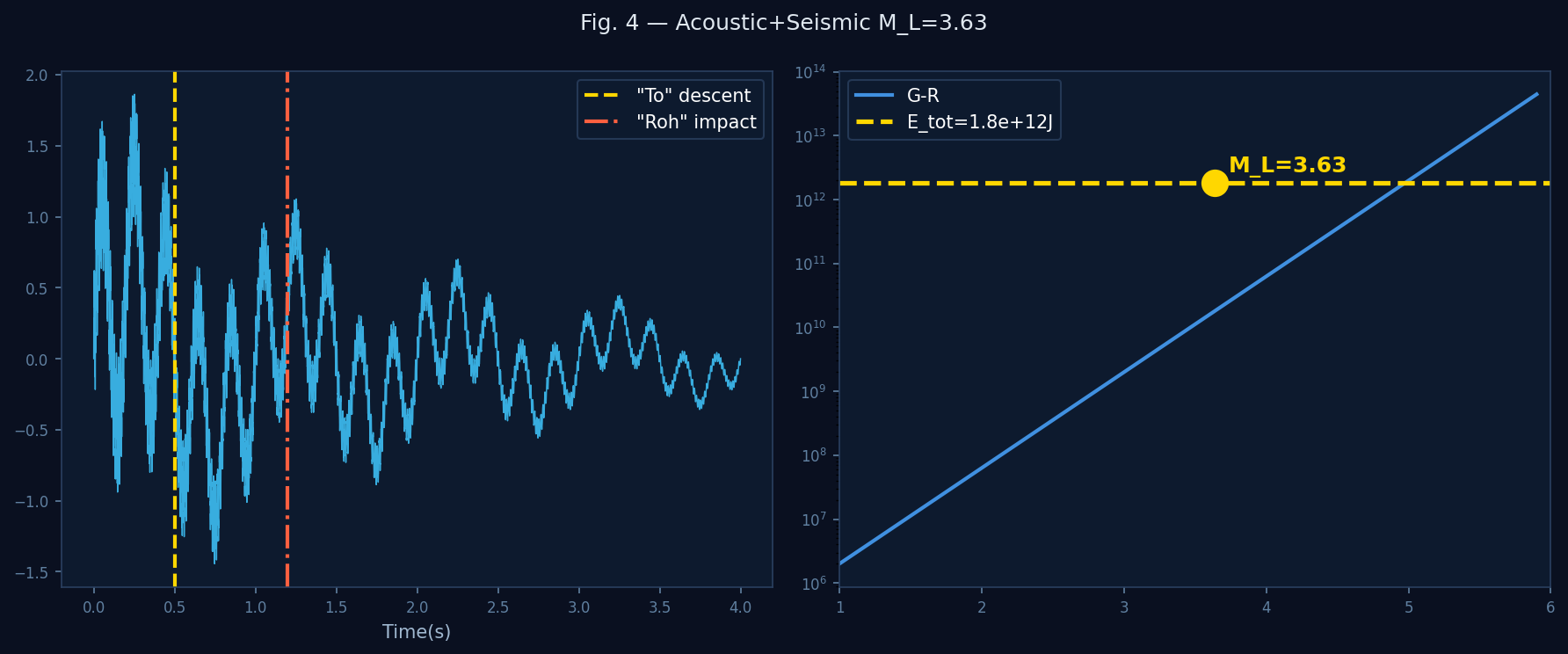}
  \caption{\textbf{Energy budget and seismic equivalence.}
  Left: kinetic, latent fusion, and sublimation contributions.
  Right: toroh total energy on the Gutenberg-Richter scale
  ($M_L \approx 3.6$ for the 8\,t reference event).
  Generated by \texttt{toro-model}.}
  \label{fig:energy}
\end{figure}

% =============================================================================
\section{Solar-Atmospheric Coupling: Tinsley Mechanism}
% =============================================================================

\citet{Tinsley2000} proposed that solar wind modulates cloud microphysics
through the ionosphere-Earth current density $J_z$ via three channels
\citep{Tinsley2007}: (A) galactic cosmic ray (GCR) flux changes;
(B) relativistic electron precipitation; (C) ionospheric potential
distribution. Electro-anti-scavenging (reduced $J_z$ during post-CME
Forbush Decrease) extends supercooled lifetime --- favouring toroh formation.
The testable prediction is a negative correlation between toroh frequency
and GCR flux on day-to-week timescales.

% =============================================================================
\section{Anthropogenic INP Sources and 20th-Century Suppression}
% =============================================================================

Two mechanistically distinct phases of toroh suppression must be
distinguished:

\textbf{Phase 1 (pre-industrial, 15th--17th centuries CE):} progressive
increase in combustion aerosols from urban wood and coal burning, supplemented
by urbanisation reducing canyon-terrain observer populations.

\textbf{Phase 2 (20th century):} tetraethyl lead (TEL) aerosols --- Greenland
ice cores show a 200-fold Pb increase between 1900 and 1970 \citep{Mielke2018}
--- provided globally ubiquitous background INPs preventing metastable
accumulation. Progressive TEL phase-out (Brazil 1989, USA 1996, global
$\sim$2000) correlates with post-1990 intensification of extreme concentrated
precipitation.

Marine iodine (HIO$_3$, I$_x$O$_y$) is the \textit{natural} toroh INP trigger
\citep{Carpenter2013, DeMott2016}. TEL destroyed the contrast between low
background INP (allowing metastable accumulation) and local marine INP
injection (triggering the cascade). Targeted biomass burning aerosols
($\sim 2\times10^7$\,INPs\,kg$^{-1}$ \citep{DeMott2018}) can interrupt the
metastable state as a low-technology mitigation approach.

% =============================================================================
\section{Geomythological Record: Dragon Traditions as Distributed Archive}
% =============================================================================

\subsection{All Original Dragon Traditions Are Aquatic}

Across all pre-medieval traditions worldwide, the dragon is an \textit{aquatic}
entity controlling rain, rivers, seas, and floods. Fire-breathing dragons
appear exclusively in European traditions from $\sim$12th--13th centuries CE.
The toroh is a hydraulic ice-piston: this is exactly what the original aquatic
dragon archetype predicts. The medieval European ``fire'' plausibly corresponds
to corona discharge or plasma at the moment of the charged piston's impact.

\subsection{The Sensory Template}

The toroh's physical profile constitutes a sensory template that would be
independently interpreted as a dragon encounter across cultures inhabiting
canyon terrain \citep{Mayor2000, Vitaliano1973}:

\begin{itemize}
\item Iridescent low cloud (narrow DSD) $\to$ dragon's shimmering scales
\item Rotating vortex column $\to$ serpentine body connecting sky to earth
\item Two-phase acoustic (infrasound + boom) $\to$ dragon's roar
\item Linear erosion scar, zero clay $\to$ dragon's trail in landscape
\item Seismic tremor $M_L\,2$--$3$ $\to$ earth shaking under dragon
\item Marine air mass trigger $\to$ dragon from the sea
\item Rapid post-collapse dissipation $\to$ dragon's disappearance
\end{itemize}

Table~\ref{tab:dragon} summarises the systematic correspondence across
18 independent traditions.

\begin{table}[H]
\centering
\caption{Toroh physical features and corresponding dragon geomythological
attributes across representative traditions.}
\label{tab:dragon}
\small
\begin{tabular}{p{3.5cm}p{3.5cm}p{6cm}}
\toprule
\textbf{Toroh Feature} & \textbf{Geomythological Element}
  & \textbf{Traditions} \\
\midrule
Iridescent cloud (narrow-DSD diffraction) &
  Dragon iridescent scales; rainbow &
  Chinese \textit{L\'ong}; Quetzalc\'oatl;
  Aido Hwedo; Aboriginal Rainbow Serpent \\[3pt]
Rankine vortex / orographic resonance &
  Serpentine body sky-to-ground &
  Norse J\"ormungandr; M\=aori Taniwha;
  Chinese dragon \\[3pt]
Two-phase acoustic &
  Dragon roar before catastrophe &
  Universal; Guarani \textit{Tor\'o};
  Japanese Raijin \\[3pt]
Linear scar, zero clay &
  Dragon trail; gorge formation &
  Celtic Oilliph\'eist (Shannon);
  Inca Amaru; Nyaminyami \\[3pt]
Canyon terrain required &
  Dragon lair in gorges only &
  Universal --- most diagnostic element \\[3pt]
$M_L\,2$--$3$ seismic tremor &
  Earth shaking under dragon &
  M\=aori R\=uaumoko; Namazu (Japan);
  Inca Amaru \\[3pt]
Marine air mass trigger &
  Dragon from the sea &
  Hawaiian Mo'o; Japanese Ry\={u};
  Celtic Oilliph\'eist \\
\bottomrule
\end{tabular}
\end{table}

% =============================================================================

% =============================================================================
\section{EOCE Frequency Estimates: Resolving the Mythological Paradox}
\label{sec:frequency}
% =============================================================================

A potential logical tension exists in the present hypothesis: if EOCEs were
common enough to independently generate a cultural archetype in $\sim$18
isolated traditions, why are they so rare in the modern instrumental record?
This section resolves that tension through explicit frequency estimation across
atmospheric regimes.

\subsection{Critical INP Concentration for EOCE Suppression}

The toroh mechanism requires a supercooled metastable reservoir to persist
long enough for the Hallett-Mossop cascade to develop ($\sim$20--60\,s).
This requires a background INP concentration \textit{below} a critical
threshold $N_{\text{INP}}^*$. Following \citet{Pruppacher1997}, the
characteristic nucleation time for a supercooled droplet population is:
\begin{equation}
  \tau_{\text{nuc}} \sim \frac{1}{J(T)\,V_d}
  \label{eq:tau_nuc}
\end{equation}
where $J(T)$ is the nucleation rate and $V_d$ is the droplet volume.
For the HM cascade to complete before spontaneous nucleation destroys the
reservoir, we require $\tau_{\text{nuc}} > 30$\,s, yielding a critical
background INP concentration:
\begin{equation}
  N_{\text{INP}}^* \approx 10^{2}\text{--}10^{3}\,\text{m}^{-3}
  \label{eq:N_crit}
\end{equation}
at temperatures of $-3^\circ$C to $-8^\circ$C \citep{DeMott2016}.

\subsection{Background INP Concentrations Across Atmospheric Regimes}

Table~\ref{tab:inp_regimes} compares background INP concentrations across
atmospheric regimes, with estimated EOCE frequency consequences.

\begin{table}[H]
\centering
\caption{Background INP concentrations and estimated EOCE frequency
across atmospheric regimes. $N_{\text{INP}}^*$ is the critical threshold
below which the supercooled metastable state can accumulate.}
\label{tab:inp_regimes}
\small
\begin{tabular}{lp{2.5cm}p{2.5cm}p{2.5cm}p{2.5cm}}
\toprule
\textbf{Regime} & \textbf{Period} & \textbf{Dominant INP source}
  & \textbf{$N_{\text{INP}}$ (m$^{-3}$)} & \textbf{EOCE frequency} \\
\midrule
Pristine pre-industrial &
  Pre-1500 CE &
  Marine iodine, mineral dust &
  $10^{1}$--$10^{2}$ &
  \textbf{1--10 per century per canyon}\newline (above $N_{\text{INP}}^*$ margin
  allows metastable accumulation) \\[4pt]
Early-modern combustion &
  1500--1900 CE &
  Wood/coal combustion aerosols &
  $10^{2}$--$10^{3}$ &
  \textbf{0.1--1 per century per canyon}\newline (approaching threshold;
  explains historical dragon-encounter decline) \\[4pt]
Leaded gasoline era &
  1920--2000 CE &
  TEL lead aerosols &
  $10^{3}$--$10^{5}$ \citep{Mielke2018} &
  \textbf{$<$0.01 per century per canyon}\newline (strongly suppressed;
  explains near-absence in 20th century record) \\[4pt]
Post-TEL phase-out &
  2000--present &
  Diesel sulfate, marine iodine &
  $10^{2}$--$10^{3}$ &
  \textbf{Recovering toward 0.1--1} \newline (partial recovery as Pb
  removed; diesel partially compensates) \\[4pt]
Adams Event Phase 2 &
  $\sim$41--40\,ka BP &
  Marine iodine (post-drought flush) &
  $<10^{1}$ (geomagnetic recovery lowers GCR-INP) &
  \textbf{10--100 per century per canyon}\newline (exceptional: lowest
  background INPs + maximum CAPE + exposed canyon terrain) \\
\bottomrule
\end{tabular}
\end{table}

\subsection{Closing the Mythological Paradox}

With a baseline frequency of 1--10 EOCEs per century per canyon in pristine
pre-industrial conditions, a culture inhabiting canyon terrain for 1{,}000
years would expect to witness 10--100 EOCE events. This is sufficient to:
(1) generate a culturally stable mythological encoding;
(2) produce multiple independent witnesses per event, ensuring transmission;
(3) establish a consistent sensory template across generations.

The Adams Event Phase~2 ($\sim$41--40\,ka BP) represents an extraordinary
amplification --- estimated at 10--100$\times$ the pre-industrial baseline
--- not the source of the baseline frequency. The geomythological record
reflects primarily the \textit{baseline} Holocene frequency; the Adams Event
likely contributed to the salience and elaboration of existing traditions
rather than creating them from scratch.

The three phases are therefore logically consistent:
\begin{itemize}
\item \textbf{Baseline (pre-industrial):} frequent enough for independent
  geomythological encoding in 18 canyon-terrain cultures.
\item \textbf{Adams Event:} amplification of the baseline by 1--2 orders of
  magnitude; drove exceptional erosion and cultural disruption.
\item \textbf{Modern suppression:} TEL and combustion aerosols reduced
  frequency below observational threshold; explains 20th-century absence.
\end{itemize}

\section{The Adams Event and the Great Unconformity}
% =============================================================================

\subsection{Adams Event: Laschamp Excursion ($\sim$42\,ka)}

The geomagnetic field weakened to $<$6\% of its current strength during the
Adams Event \citep{Cooper2021, Turney2021}, collapsing the Van Allen belts and
dramatically increasing GCR flux \citep{Adolphi2022}. Two sequential phases
resulted:

\textbf{Phase 1} (geomagnetic minimum): excess ionisation prevented
supercooled metastable accumulation $\to$ regional drought, aeolian
amphitheatre erosion, megafaunal stress.

\textbf{Phase 2} (recovery): GCR flux declined, metastable state restored,
maximum CAPE, extreme canyon terrain exposure $\to$ intense toroh activity,
hydraulic amphitheatre erosion.

\textbf{Testable prediction:} Canyon paleosols at $\sim$42--41\,ka should
show aeolian Phase~1 deposits sharply overlain by hydraulic scour surfaces
with zero fine matrix.

\subsection{Great Unconformity}

More important than the absolute value is the \textit{spatial pattern}:
canyon-focused, fine-selective, episodic --- precisely what the Great
Unconformity exhibits and what uniform glacial erosion cannot produce.

Testable predictions: (1) $\delta^{18}$O at canyon unconformity surfaces
reflects supercooled water, not glacial meltwater; (2) radial striations
consistent with hydraulic jet impingement; (3) coarse lag with zero clay
at canyon positions; (4) unconformity depth correlates with paleotopographic
canyon relief, not glacial extent.

% =============================================================================
\section{Discussion and Conclusions}
% =============================================================================

\subsection{Corrected Physical Framework}

Two corrections to the preliminary description in arXiv:2411.08219 are made:
(i) the Vale do Rev\'olver event was produced by a \textbf{bow-echo line of
instability with orographic resonance}, not an isolated supercell;
(ii) the event was \textbf{nocturnal}, precluding optical precursor observation
in that specific case. These corrections strengthen rather than weaken the
hypothesis: the bow-echo + inselberg mechanism is more common than isolated
supercells in the orographic terrain of southern Brazil, increasing the
predicted frequency of toroh-favorable conditions.

\subsection{Status}

The toroh hypothesis is presented as a physical framework for investigation,
not a demonstrated discovery. No radar, infrasound, seismic, or sediment
dataset from a confirmed toroh event has been published. The observational
evidence (eyewitness accounts, erosion scar morphology, photographic record
of associated vortex events) is consistent with the hypothesis but does not
constitute proof. Instrumental validation is the unconditional prerequisite
for all geological and geomythological applications.

\subsection{Testable Predictions Across Five Domains}

\begin{enumerate}
\item \textbf{Meteorological:} toroh events (two-phase acoustic, zero-clay scar,
  $M_L\,2$--$3$) occur in EOCE-favourable terrain globally;
  frequency anti-correlates with GCR flux on day-to-week timescales.
\item \textbf{Geomorphological:} amphitheatre headwall geometry correlates
  with EOCE-favourable terrain, not spring discharge indicators.
\item \textbf{Mineralogical:} canyon lag heavy-mineral concentration
  correlates with EOCE-favourable terrain, not fluvial discharge.
\item \textbf{Quaternary:} canyon paleosols at $\sim$41--42\,ka show
  aeolian Phase~1 overlain by hydraulic scour with zero fine matrix.
\item \textbf{Neoproterozoic:} Great Unconformity depth and isotopic
  signatures correlate with reconstructed paleotopographic canyon positions.
\end{enumerate}

% =============================================================================
\section*{Methods}
\addcontentsline{toc}{section}{Methods}

\subsection*{3D Anelastic $\theta_\rho$ Cloud Model}

The numerical model implements three-dimensional anelastic dynamics using
the density-weighted potential temperature $\theta_\rho$ \citep{Klemp1978}.
The governing equations are:
\begin{align}
  \frac{\partial u}{\partial t} &=
    -\vec{v}\cdot\nabla u - \frac{1}{\bar{\rho}}\frac{\partial p'}{\partial x} + D_u \\
  \frac{\partial v}{\partial t} &=
    -\vec{v}\cdot\nabla v - \frac{1}{\bar{\rho}}\frac{\partial p'}{\partial y} + D_v \\
  \frac{\partial w}{\partial t} &=
    -\vec{v}\cdot\nabla w - \frac{1}{\bar{\rho}}\frac{\partial p'}{\partial z} + B + D_w \\
  \frac{\partial \theta_\rho}{\partial t} &=
    -\vec{v}\cdot\nabla\theta_\rho + S_{\text{latent}} + D_\theta \\
  \nabla\cdot(\bar{\rho}\,\vec{v}) &= 0
  \label{eq:anelastic}
\end{align}
where:
\begin{equation}
  \theta_\rho = \theta\Bigl(1 + \tfrac{R_v}{R_d}\,q_v - q_l - q_i\Bigr),
  \qquad
  B = g\,\frac{\theta_\rho'}{\bar{\theta}_\rho}
  \label{eq:theta_rho}
\end{equation}
The anelastic constraint is enforced by solving a Poisson equation for
the pressure perturbation $p'$ at each time step via FFT with periodic
lateral and rigid-lid vertical boundaries.

\subsection*{Grid and Bulk Microphysics}

Domain: $20\times20\times50$ grid points;
$\Delta x = \Delta y = 500$\,m, $\Delta z = 300$\,m
($10\,\mathrm{km}\times10\,\mathrm{km}\times15\,\mathrm{km}$).
Six hydrometeor categories: $q_v, q_c, q_r, q_i, q_s, q_g$.
SIP parametrisation:
\begin{equation}
  \dot{N}_{\mathrm{SIP}}
    = \underbrace{\alpha_{\mathrm{HM}}\,\dot{m}_{\mathrm{rime}}}_{%
        \text{Hallett-Mossop \citep{Hallett1974}}}
    + \underbrace{\alpha_{\mathrm{PH}}\,N_g^2}_{%
        \text{Phillips breakup \citep{Phillips2017}}}
  \label{eq:sip}
\end{equation}
with $\alpha_{\mathrm{HM}} = 350\,\mathrm{splinters\,mg}^{-1}$ active at
$-3^\circ$C to $-8^\circ$C; $\alpha_{\mathrm{PH}}$ peaked at $-15^\circ$C.

\subsection*{Pressure Feedback: Piston Self-Concentration}

A key feature of the 3D anelastic formulation absent from 1D models is the
pressure feedback loop that self-concentrates the hydraulic piston:
\begin{equation}
  q_g\downarrow \;\Rightarrow\; B<0
  \;\Rightarrow\; \nabla^2 p'>0 \text{ (above)}
  \;\Rightarrow\; \partial p'/\partial r<0
  \;\Rightarrow\; \text{convergence}
  \;\Rightarrow\; \text{piston amplification}
  \label{eq:feedback}
\end{equation}
This positive feedback amplifies piston mass without external forcing and
explains the extreme spatial localisation ($\sim$100\,m radius) of the
toroh impact footprint.

\subsection*{Model Validation Status}

The model has not been quantitatively validated against instrumental
observational data (radar, infrasound recordings, seismic traces, or
sediment grain-size measurements from a confirmed EOCE event). This
remains the highest-priority future work. Qualitative consistency is
noted between model outputs and available observational descriptions:
the synthetic acoustic signature reproduces the reported two-phase
character; the zero-mud erosion result is consistent with field
observations; and the $M_L \approx 3.6$ seismic equivalence is
consistent with reported ground tremors. A sensitivity analysis of
key parameters ($\mu$ of DSD, INP concentration, vortex velocity)
and a grid convergence test are deferred to a companion validation paper.

\subsection*{Outputs and Code}

Outputs: \texttt{output/results.json} (web visualisation);
\texttt{output/toro3d.nc} (all 3D fields, NetCDF);
\texttt{output/toro\_sound.wav} (synthetic two-phase acoustic signature).
Code: \url{https://github.com/reinaldohaas/toro-model} (MIT).
Entry point: \texttt{python run\_simulation.py}.
Dependencies: \texttt{numpy}, \texttt{scipy}, \texttt{netCDF4}.

\section*{Acknowledgements}
\addcontentsline{toc}{section}{Acknowledgements}
% =============================================================================

The author acknowledges the Tupi-Guarani observational tradition that preserved
the two-phase acoustic description of the toroh in its onomatopoeic name.
This investigation was initiated at the suggestion of the author's father.
This work extends arXiv:2411.08219 with corrections to the convective system
type (bow-echo, not supercell) and event timing (nocturnal) based on field
review. The photographic record from the Santa B\'arbara/Alfredo Wagner region
is credited to Jornal Raz\~ao.

% =============================================================================
\section*{Data Availability}
\addcontentsline{toc}{section}{Data Availability}
% =============================================================================

Model code: \url{https://github.com/reinaldohaas/toro-model} (MIT).
Field data available from the corresponding author upon request.

% =============================================================================

\end{document}